\documentclass[twocolumn,preprintnumbers,amsmath,amssymb,prl]{revtex4-1}

\usepackage{amsmath,amssymb,,epsfig}

\usepackage{dcolumn}

\def\be{\begin{eqnarray}}
\def\ee{\end{eqnarray}}
\def\bea{\begin{eqnarray}}
\def\eea{\end{eqnarray}}

\newcommand{\Tr}{\mbox{Tr}}

\def\S{{\mathbb S}}
\def\Pf{\text{Pf}}

\def\e{\epsilon}

\begin{document}

\preprint{}

\title{
A ``Lagrangian'' for a non-Lagrangian theory
}

\author{Abhijit Gadde,$^1$ Shlomo S. Razamat,$^{2,3}$ and Brian Willett$^1$}

\affiliation{$^1$Institute for Advanced Study, Princeton, NJ 08540, USA\\
$^2$NHETC, Rutgers, Piscataway, NJ 08854, USA \\
$^3$Department of Physics, Technion, Haifa 32000, Israel\\
}

\date{\today}

\begin{abstract}

We suggest an ${\cal N}=1$ Lagrangian flowing in the infra-red to the  ${\cal N}=2$ rank one  superconformal field theory with $E_6$ flavor symmetry. We utilize this description to compute several supersymmetric partition functions.

\end{abstract}


\maketitle


\noindent{\bf Introduction:\;\;\;}
The exploration of the landscape of supersymmetric quantum field theories in four dimensions inevitably leads us to  consider strongly coupled isolated theories {\it i.e.} interacting theories with no conformal manifold. 
In the case of ${\cal N}=1$ supersymmetry, such isolated theories can have an interpretation as the infra-red fixed point of  renormalization group flow. Naturally, this is due to a multitude of relevant deformations admitted by the ${\cal N}=1$ supersymmetric Lagrangians. Theories with ${\cal N}=2$ supersymmetric Lagrangians on the other hand, do not admit any ${\cal N}=2$ relevant superpotential deformations except for the mass terms. 
This has given ${\cal N}=2$ isolated theories a somewhat mysterious status of being non-Lagrangian. In this note, we point out a way around this obstacle by discussing an ${\cal N}=1$ RG flow whose supersymmetry is enhanced to ${\cal N}=2$ at the fixed point. 

Our strategy is to look for special points on the conformal manifold of an ${\cal N}=2$ theory, where the symmetry commuting with the ${\cal N}=1$ supersymmetry is enhanced. Although the RG flow obtained by gauging that symmetry with an ${\cal N}=1$ vector multiplet may not always end up in an ${\cal N}=2$ isolated theory, we argue that it \emph{does} happen at least in certain cases. 

In this note we  discuss an ${\cal N}=1$ Lagrangian description for the ${\cal N}=2$ superconformal theory with $E_6$ flavor symmetry~\cite{Minahan:1996fg}.
The inspiration for our construction is the structure of the exact expression for the superconformal index of this theory  which can be obtained using, a priori, mathematical acrobatics~\cite{Gadde:2010te}. 
The Lagrangian we will derive, although having certain shortcomings, will be  illustrated to be  quite useful for performing a variety of explicit computations.

\noindent{\bf Argyres-Seiberg duality:\;\;\;}
Let us start by discussing the basics of the Argyres-Seiberg duality~\cite{Argyres:2007cn} which is essential for our construction. On side A of the duality we have 
$SU(3)$ ${\cal N}=2$ SQCD with $N_f=6$ fundamental hypermultiplet flavors. On side B of the duality we have the ${\cal N}=2$ Minahan-Nemeschansky $E_6$ SCFT~\cite{Minahan:1996fg}, with an $SU(2)_g$ subgroup of the $E_6$ flavor symmetry gauged with an addition of a single hypermultiplet, $(Q_B,\widetilde Q_B)$, in the fundamental representation of the gauge group. The global symmetry of side A is $U(6)$ which we choose to decompose into $SU(3)_a\times SU(3)_b\times U(1)_s\times U(1)_y$. The charges of all the fields are given in the first part of Table~\ref{tab:charges}. On side B the symmetries are identified as follows. 
The $E_6$ symmetry has an $SU(3)_a\times SU(3)_b\times SU(3)_c$ maximal subgroup. We gauge an $SU(2)_g$ subgroup of $SU(3)_c \supseteq (SU(2)_g\times U(1)_y)/{\mathbb Z}_2$. The $(Q_B,\widetilde Q_B)$ hypermultiplet is in fundamental representation of  $SU(2)_g$ and the half-hypers are rotated by $U(1)_s$.  The two descriptions are conformal and have a single exactly marginal coupling with the relation,
\be\label{coups}
\tau_B=\frac1{1-\tau_A}\,,\qquad\; \tau=\frac\theta\pi+\frac{8\pi i}{g^2}\,. 
\ee In particular the strongly-coupled cusp on side A, $\tau_A=1$, is mapped to zero coupling of the $SU(2)_g$ gauge group on side $B$. We also note that the ${\cal N}=2$ supersymmetry in four dimensions has $SU(2)_R\times U(1)_r$ R-symmetry with the hypermultiplets being doublets of $SU(2)_R$ and having no charge under $U(1)_r$, and  the adjoint chiral in the ${\cal N}=2$ vector being singlet of $SU(2)_R$ but having $U(1)_r$ charge $-1$.

\noindent{\bf Deformation of side B:\;\;\;} Let us now consider a deformation of side B of the duality.  This theory has a superpotential dictated by ${\cal N}=2$ supersymmetry involving the term,
\be
W=Q_B\Phi_B \widetilde Q_B+\Phi_B \hat\mu\,,
\ee with $\Phi_B$ being the adjoint chiral in the $SU(2)_g$ vector multiplet and $\hat\mu$ being the moment map operator in the $E_6$ SCFT for $SU(2)_g$ symmetry. We consider a  deformation which removes the first term from the superpotential.  
The supersymmetry is broken to ${\cal N}=1$. Note however that the only symmetry broken by the deformation is the non-abelian part of $SU(2)_R$ R-symmetry. All the abelian generators are  preserved. In particular we can define the symmetry $U(1)_t$ generated by $R^3+r$ which is a flavor, non-R-symmetry, of the deformed theory.  The R-symmetry of the remaining ${\cal N}=1$ supersymmetry can be taken to be $-2r$ and we will denote it as $U(1)_{\hat r}$.

An interesting fact about the deformed theory is that the $U(1)_s$ symmetry rotating the half-hypermultiplets enhances to $SU(2)_s$. The $SU(2)$ enhancement was broken in the ${\cal N}=2$ theory by the superpotential term we now removed. 
Note that the deformation we introduce is classically marginal and
quantum mechanically has to be marginal and/or marginally irrelevant~\cite{Green:2010da}, meaning that it flows back to the conformal manifold.
The only point on the conformal manifold with $U(1)_s$ enhanced to $SU(2)_s$ is the free point for the $SU(2)_g$ gauge theory and that would be the expected IR fixed point of our deformation. We will soon come to discuss the deformation on side A  of the duality.

\noindent{\bf Gauging $SU(2)_s$/``un-gauging'' $SU(2)_g$:\;\;\;}  We can consider an ${\cal N}=1$ gauging of the $SU(2)_s$
symmetry of the deformed theory. (Note that even at the free point gauging $SU(2)_s$ is not consistent with ${\cal N}=2$ supersymmetry.)  We choose to gauge this symmetry after adding two chiral fields in the fundamental representation of $SU(2)_s$.   We will denote the components with $\pm1$ charge under the Cartan $U(1)_s$ as $q^{1,2}$ and ${\widetilde q}^{1,2}$ The symmetry rotating $q^{1,2}\, ({\widetilde q}^{1,2})$ will be denoted $SU(2)_w$.
 The $SU(2)_s$ ${\cal N}=1$ gauge theory thus obtained has $N_f=2$ with the four fundamental chirals  given by $(q^{1,2},{\widetilde q}^{1,2})$, $Q_B$, and $\widetilde Q_B$. The flavor symmetry of this part of the theory is classically $SU(4)\supseteq SU(2)_g\times SU(2)_w\times U(1)_t$. Quantum mechanically  this $SU(4)$ symmetry is broken to $Sp(4)$~\cite{Seiberg:1994bz}. The breaking occurs because of the quantum constraint,
\be
\Pf\; M =\Lambda^4\,,
\ee 
where $\Lambda$ is the dynamical scale and $M$ is the antisymmetric matrix of the 
mesons/baryons built from the quarks $((q^{1,2},{\widetilde q}^{1,2}), Q_B,\widetilde Q_B)$.  In all vacua the gauge group $SU(2)_s$ is Higgsed because some component of $M$ gets a non zero vacuum expectation value (vev). 
There are vacua on the moduli space  where mesonic operators obtaining a vev are also charged under $SU(2)_g$ and thus
 the symmetry $SU(4)\supseteq SU(2)_g\times SU(2)_w\times U(1)_t$ is broken to a diagonal $SU(2)$, which by abuse of notation we will also denote as $SU(2)_w$, and the $U(1)_t$ remains. 
The $SU(2)_g$ gauge symmetry in such vacua is thus also Higgsed. 
At the level of the
supersymmetric index the above translates into a neat observation that the index of $SU(2)$ $N_f=2$ theory behaves as a delta-function implementing the breaking of $SU(4)$ to $Sp(4)$~\cite{Spiridonov:2014cxa} which gives a physical meaning to the Spridonov-Warnaar inversion formula~\cite{SpiridonovInver} used to obtain the index of $E_6$ SCFT in~\cite{Gadde:2010te}.

The theory in the IR  consists of two baryonic operators, $B_T=\e\cdot q_1\widetilde q_2$ and $B_{T'}=\e \cdot Q_B\widetilde Q_B$ (where $\e$ contracts the $SU(2)_g$ indices) coupled to $\Phi_B$ and the $E_6$ SCFT through a superpotential.
We can remove these fields by adding new chiral fields $\mu$, $T$ and $T'$ with a a quadratic superpotential,
\be
\mu \Phi_B+T B_T+T' B_{T'}\,.
\ee  We conclude that the procedure 
of deforming theory B, gauging the $SU(2)_s$ enhanced symmetry, and adding appropriate singlet fields and superpotential terms gives us the $E_6$ SCFT by itself. Note that in the construction many symmetries are not manifest.
The ${\cal N}=1$  supersymmetry is expected to enhance to ${\cal N}=2$ in the IR and the flavor symmetry $SU(2)\times SU(6)$ to enhance to $E_6$.

\noindent{\bf Deformation and gauging on side A:\;\;\;}
Let us discuss now the procedure in the Lagrangian duality frame. Since the two duality frames are exactly 
equivalent (this is a conformal duality), we should be able to retrace all our steps on side A of the duality.
First, the superpotential deformation we introduce, $-Q_B\Phi_B \widetilde Q_B$, breaks no (${\cal N}=2$) flavor symmetries and the natural candidate on side A for it to map to
is $- C(\tau_A)\, Q^i_A\Phi_A {\widetilde Q}^i_A$ with $C(\tau_A)$ a proper normalization factor.
In fact the normalization factor $C(\tau_A)$  is plausibly infinite as the following argument suggests. Note that as we discussed above theory B after deformation will flow to the zero coupling locus of the conformal manifold. That point in theory A corresponds to infinite gauge coupling~\eqref{coups} . 
The enhancement of $U(1)_s$ to $SU(2)_s$ should occur in the limit of infinite coupling of superpotential of theory A. In that limit we can gauge $SU(2)_s$ symmetry and straightforwardly repeat all the steps performed on side B.

This procedure  provides us with a Lagrangian description of the $E_6$ SCFT. The Lagrangian is not conformal, does not have the manifest non abelian structure of the flavor symmetry,  has only ${\cal N}=1$ supersymmetry, and only makes sense in the limit of one of the superpotential couplings taken to infinity. 

To summarize, the Lagrangian we propose is the $SU(3)$ ${\cal N}=2$ SQCD with $N_f=6$ with 
the superpotential deformed to the $SU(2)_s$ enhanced point and the $SU(2)_s$ symmetry subsequently gauged with an addition of two fundamental chiral fields and three singlet chirals. The charges of the fields are summarized in Table~\ref{tab:charges} and one should include any superpotential term consistent with those. 

\begin{table*}
\caption{\label{tab:charges} The field content of the Lagrangian description of the $E_6$ SCFT.
We denote the charges under $U(1)_s$ with the enhancement to $SU(2)_s$ not manifest in this description. 
The upper part of the table is the $SU(3)$ SCFT. The middle part has the chirals added when gauging $SU(2)_s$ and the bottom part has the gauge singlet fields.}
\begin{ruledtabular}
\begin{tabular}{c|c|c||c|c|c|c|c|c}
 Field&$SU(3)$&$U(1)_s$&$SU(3)_a$&$SU(3)_b$&$U(1)_{\hat r}$&$U(1)_t$&$U(1)_y$&$SU(2)_w$\\ \hline
 $\Phi_A$&${\bf 8}$&$0$ &${\bf 1}$&${\bf 1}$& $2$& $-1$& $0$&${\bf 1}$ \\
$Q^1_A$&${\bf 3}$&$\frac13$&${\bf 3}$&${\bf 1}$&$0$&$\frac12$&$-1$&${\bf 1}$\\ 
${\widetilde Q}^1_A$&$\bar{\bf 3}$&$-\frac13$&$\bar{\bf 3}$&${\bf 1}$&$0$&$\frac12$&$1$&${\bf 1}$\\ 
$Q^2_A$&$\bar{\bf 3}$&$-\frac13$&${\bf 1}$&${\bf 3}$&$0$&$\frac12$&$-1$&${\bf 1}$\\ 
${\widetilde Q}^2_A$&${\bf 3}$&$\frac13$&${\bf 1}$&$\bar {\bf 3}$&$0$&$\frac12$&$1$&${\bf 1}$\\ \hline
$q$&${\bf 1}$&$1$&${\bf 1}$&${\bf 1}$&$0$&$-\frac12$&$0$&${\bf 2}$\\ 
$\widetilde q$&${\bf 1}$&$-1$&${\bf 1}$&${\bf 1}$&$0$&$-\frac12$&$0$&${\bf 2}$\\  \hline
$T$&${\bf 1}$&$0$&${\bf 1}$&${\bf 1}$&$2$&$1$&$0$&${\bf 1}$\\
$T'$&${\bf 1}$&$0$&${\bf 1}$&${\bf 1}$&$2$&$-1$&$0$&${\bf 1}$\\
$\mu$&${\bf 1}$&$0$&${\bf 1}$&${\bf 1}$&$0$&$1$&$0$&${\bf 3}$\\
\end{tabular}
\end{ruledtabular}
\end{table*}

\noindent{\bf a-maximization:\;\quad\;} Since the Lagrangian we have obtained hinges on $U(1)_s$ symmetry being enhanced to $SU(2)_s$ for a fine-tuned value of the superpotential coupling, which moreover is infinite, one can wonder how useful it is. Nevertheless, there are quite a few computations which can be performed relying only on symmetries, matter content, and gauge interactions of the Lagrangian. As a simple example we can use the Lagrangian to compute the $a$ and $c$ anomalies of the $E_6$ SCFT  utilizing a-maximization~\cite{Intriligator:2003jj}. The only abelian symmetry we have which can be admixed to the R-symmetry is $U(1)_t$ and thus parametrizing the IR R-charge as 
$r_{IR}=\hat r+S \, t$, we obtain that using the matter content and the gauge interactions detailed in Table~\ref{tab:charges} the trial $a_{trial}(S)$  and $c_{trial}(S)$ anomalies are given by,
\be
&&a_{trial}(S)=-\frac3{32}(3S^3+27S^2-88S+44)\,,\\
&&c_{trial}(S)=-\frac1{32}(9S^3+81S^2-242S+88)\,.\nonumber
\ee The trial a-anomaly is maximized for $S=\frac43$ which gives us,
\be
a_{IR}=\frac{41}{24}\,,\qquad\; c_{IR}=\frac{13}6\,.
\ee These are the correct anomalies of the $E_6$ SCFT.

\noindent{\bf Partition functions:\;\;\;} Another natural check of the Lagrangian  would involve the computation of the supersymmetric index, and that check was in fact performed in~\cite{Gadde:2010te} though not given the physical meaning we advocate in this note.  In addition to the index there are other partition functions one can straightforwardly compute given a Lagrangian. These include the lens space 
index~\cite{Benini:2011nc} and the $\S^2\times {\mathbb T}^2$~\cite{Benini:2015noa} partitions functions. These partition functions capture also information about non-local objects which are not counted by the supersymmetric index. 

Let us discuss here a particular case of the lens space index, $\S^3/{\mathbb Z}_{\frak r}\times \S^1$ partition function, {\it i.e.} the limit of infinite $\frak r$ when it becomes  the supersymmetric index of the dimensionally reduced theory.
The $E_6$ SCFT reduced to three dimensions has a mirror Lagrangian description as a quiver theory~\cite{Benini:2010uu}. We can compute supersymmetric partition functions in the mirror frame and compare them to the computation performed in our Lagrangian reduced to three dimensions. 
The equality of $\S^3$ partition functions is guaranteed by the fact that the correct index in four dimensions, $\S^3\times \S^1$ partition function,  is produced by the Lagrangian. However the three dimensional index, the $\S^2\times \S^1$ partition function, will be an independent check.
The ${\cal N}=4$ index in three dimensions is given by,
\be
{\cal I}=\Tr(-1)^F {\frak q}^{j+\frac12(R_H+R_C)}{\frak t}^{R_H-R_C}\,,
\ee where we use notations of~\cite{Razamat:2014pta}. Taking the matter content of the dimensionally reduced theory and the map of the R-charges $(R_H,R_C)=(R,-r)$, we can compute the index of the dimensionally reduced theory without any $\S^2$ fluxes for global symmetries to be,
\be
{\cal I}=1+{\bf 78} {\frak q}^{\frac12} {\frak t}-(1+{\bf 78}) {\frak q}+{\bf 2430} {\frak q}{\frak t}^2+\cdots\,.
\ee  The spectrum falls into $E_6$ irreps and is consistent with the index computed in the mirror description. 
Computing the index is absolutely standard but one should be careful with two points. First, the contour of integration for $SU(2)_s$ has to separate sequences of poles converging to zero and infinity  in the $SU(2)_s$ fugacity coming from $q$ and $\widetilde q$~\cite{Gadde:2010te}. Second, one has to be careful with global properties of the gauge and flavor groups. For example, $SU(3)_c\supseteq (SU(2)_w\times U(1)_r)/{\mathbb  Z}_2$ and the ${\mathbb Z}_2$ implies that $SU(2)_w$ and $U(1)_r$ fluxes can be simultaneously half-integer. Similarly, since $U(1)_s$ charges of the $SU(3)$ gauge theory are fractional, taking (half-) integer fluxes for $SU(2)_s$  which are not multiples of three, the fluxes for $SU(3)$ take values in an appropriately shifted lattice.

A simple version of the index is the Coulomb limit~\cite{Razamat:2014pta}, taking ${\frak t},{\frak q}\to0$ while keeping the ratio $\frac{{\frak q}^{\frac12}}{{\frak t}}=x$ fixed. In this limit the dependence on fugacities under which Higgs branch operators are charged drops out but the dependence on the fluxes remains. Given that the index in the Coulomb limit of the $SU(3)$ SYM is 
${\cal I}_C(m^a_1,m^a_2,m^b_1,m^b_2,m^s,m^y)$ (where $m$s are the $\S^2$ fluxes for global symmetries), gauging the $SU(2)_s$ symmetry and adding the fields in Table~\ref{tab:charges} it can be shown that the Coulomb index of the $E_6$ SCFT is,
\be
&&{\cal I}^{E_6}_C(m^a_1,m^a_2,m^b_1,m^b_2,m^w=m^s,m^y) = x^{|m^s|}\times\\
&&\left((1+x){\cal I}_C(;m^s;)-x{\cal I}_C(;|m^s|-1;)-{\cal I}_C(;|m^s|+1;)\right)\,. \nonumber
\ee This expression has to be invariant under the Weyl symmetry of $E_6$ and one can check that indeed it is. Some details 
of this computation are provided in accompanying Mathematica notebook.

Next we consider the ${\cal N}=1$ supersymmetric $\S^2 \times {\mathbb T}^2$ index~\cite{Benini:2015noa}.  Here one works in a background with an R-symmetry gauge field with unit flux on the $\S^2$, and in order for the fields to live in well-defined bundles, one must pick an R-symmetry under which they have integer charges~\cite{Closset:2013vra}. Our choice of $U(1)_{\hat r}$ satisfies this property.  In particular, this R-symmetry preserves all the flavor symmetry of the ${\cal N}=2$ theory, and so we can check that the index of the above Lagrangian theory demonstrates the expected $E_6$ property.  
The $\S^2 \times {\mathbb T}^2$ index is a function of the complex structure of the torus, $q=e^{2 \pi i \tau}$, as well as holonomies for flat background gauge fields on the torus coupled to each flavor symmetry, which are organized into a complex fugacity, $z=e^{2 \pi i (A_1 + \tau A_2)}$.  For the Lagrangian  described above, the relevant fugacities are $t$ for the $U(1)_t$  and $\alpha_a$, $a=1,...,6$, for the flavor symmetry which we expect to enhance to $E_6$.  The index admits a double series expansion in $q$ and $t$, and we can observe that the dependence on $\alpha_a$ at each order organizes into $E_6$ representations.  The computation here is again mostly straightforward, with similar subtleties as in the three dimensional index discussion. 
In the attached Mathematica notebook we provide with the full expression of this index and here we present the result for the first few terms in the expansion,
\be 
&{\cal I}_{\S^2 \times {\mathbb T}^2} = t^{11/2}(1\;+\; t \;{\bf 78} \;+\; t^2 \;{\bf 2430} \;+\;  t^3 \;{\bf 43758} + \cdots) \nonumber \\
& + q \; t^{11/2} ( 1\;+\;{\bf 78}  \;+\; 
 t (1 \;+\; 2 \; {\bf 78} \;+\;  {\bf 2430} \;+\;  {\bf 2925}) + \nonumber \\
& \;t^2 ({\bf 78} \;+\; 
    2 \;{\bf 2430} \;+\; {\bf 2925} \;+\;{\bf 43758} \;+\;  {\bf 105600}) \;+\; \cdots) \nonumber \\
& + q^2\; t^{11/2} ( 3 \;+\; 2 \;{\bf 78}  \;+\; {\bf 650} \;+\; {\bf 2430} \;+ \cdots) + \cdots  .\nonumber \\ 
&
\ee
An interesting feature is that the $q \rightarrow 0$ limit of this index  coincides, up to an overall factor, with the Hall-Littlewood limit of the superconformal index of the $E_6$ SCFT~\cite{Benvenuti:2010pq}.  Thus the $\S^2 \times {\mathbb T}^2$ index of ${\cal N}=2$ theories with our choice of R-symmetry appears to be an elliptic generalization of the Hall-Littlewood index/Higgs branch Hilbert series.

\noindent{\bf Generalizations:\;\;\;} 
The derivation of a Lagrangian for the $E_6$ SCFT suggested here is based on giving up the manifest ${\cal N}=2$ supersymmetry and
manipulating enhanced symmetry  of special points of the parameter space. We can apply the same  procedure 
to an infinite sequence of ${\cal N}=2$ SCFTs. Take class ${\cal S}$ $A_{N-1}$  theory corresponding to sphere with two maximal and $N-1$ minimal punctures. This theory has two dual descriptions, one as a Lagrangian linear quiver and second as the $T_N$ SCFT
coupled to superconformal tail terminating with $SU(2)$ gauge group~ \cite{Gaiotto:2009we}. We can apply our procedure to ``un-gauge'' this $SU(2)$ which transforms two minimal punctures into an $SU(2)\times U(1)$ $L$-shaped puncture. Turning on then a general vev on the baryonic branch one removes the $N-3$ remaining minimal punctures ending with a sphere 
with two maximal and one $L$-shaped puncture, which is a non-trivial SCFT. Our procedure can be thought of as providing a Lagrangian for these models, using which, {\it e.g.}, the supersymmetric partition functions can be computed. 

It would be very interesting to generalize this type of 
construction to other SCFTs in the ${\cal N}=2$ theory space~\cite{Gaiotto:2009we}. 
Although the Lagrangian we obtain requires  singular superpotentials we have illustrated that it is still very useful to obtain non-trivial results about the $E_6$ SCFT. It would be extremely interesting to perform further checks of our proposal as well as extract new information about  these strongly coupled field theories.

\noindent{\bf Acknowledgments}:~
We would like to thank C.~Beem and L.~Rastelli for useful comments. 
AG is supported by the Raymond and Beverly Sackler Foundation Fund and the NSF grant PHY-1314311. SSR was partially supported by Research in Theoretical High Energy Physics grant DOE-SC00010008. BW is supported by DOE Grant de-sc0009988 and the Roger Dashen Membership.

\end{document}